\documentclass[onecolumn,floatfix,aps,superscriptaddress]{revtex4}
\usepackage{graphicx}
\usepackage{amsmath}
\usepackage[colorlinks=true, citecolor=blue, urlcolor=blue, linkcolor = blue]{hyperref}
\usepackage{amssymb}
\usepackage{bm}
\usepackage{color}
\usepackage{bookmark}
\usepackage{tabularx}
\usepackage{mathtools}
\usepackage{microtype}
\usepackage{cancel}
\usepackage{relsize}

\usepackage{xcolor}
\begin{document}
\title{Accelerated magnetosonic lump wave solutions by orbiting charged space debris }
\author{S. P. Acharya
\footnote{Electronic mail: siba.acharya@saha.ac.in and siba.acharya39@gmail.com}}
\affiliation{Saha Institute of Nuclear Physics, a Constituent Institute of Homi Bhabha National Institute (HBNI), 1/AF Bidhannagar, Kolkata-700064 (India)}
\author{A. Mukherjee
\footnote{Electronic mail: abhikmukherjeesinp15@gmail.com}}
\affiliation{Physics and Applied Mathematics Unit, Indian Statistical Institute, Kolkata, India}
\author{M. S. Janaki
\footnote{Electronic mail: ms.janaki@saha.ac.in}}
\affiliation{Saha Institute of Nuclear Physics, a Constituent Institute of Homi Bhabha National Institute (HBNI), 1/AF Bidhannagar, Kolkata-700064 (India)}
%\date{}
\begin{abstract}
The excitations of nonlinear magnetosonic  lump waves induced by orbiting charged space debris particles in the Low Earth Orbital (LEO) plasma region are investigated in presence of the ambient magnetic field. These nonlinear waves are found to be governed by the forced Kadomtsev-Petviashvili (KP) type model equation, where the forcing term signifies the source current generated by different possible motions of charged space debris particles in the LEO plasma region. Different analytic lump wave solutions that are stable for both slow and fast magnetosonic waves in presence of charged space debris objects are found for the first time. The dynamics of exact pinned accelerated magnetosonic lump waves is explored in detail. Approximate magnetosonic lump wave solutions with time-dependent amplitudes and velocities are analyzed through perturbation methods for different types of localized space debris functions; yielding approximate pinned accelerated magnetosonic lump wave solutions. These new results may pave new directions in this field of research. 
\end{abstract}
\maketitle
\section{Introduction} \label{Intro}
Upsurge in the  research endeavors encompassing the dynamics of space debris objects in the near-earth atmosphere has been gaining significant attention by scientific community from the middle of last century. Space debris objects \cite{Klinkrad} include different kinds of dead satellites, destroyed spacecrafts, meteoroids, other inactive materials in space resulting from many natural phenomena, and are being levitated in extraterrestrial regions  especially in the near-earth space. These are also  referred to as space junk. The space debris objects are substantially found in the Low Earth Orbital (LEO) \cite{Sampaio} and Geosynchronous Earth Orbital (GEO) regions. Also, their number is continuously being increased nowadays due to increasing number of artificial space missions resulting in dead satellites,  destroyed spacecrafts etc., and many natural hazards occurring in space. These debris objects become charged in a plasma medium because of different mechanisms such as photo-emission, electron and ion collection, secondary electron emission \cite{Horanyi} etc. These charged debris objects are  of varying sizes ranging from as small as microns to as big as centimeters.  In certain conditions, these debris objects  move with different velocities; causing significant harm to running spacecrafts \cite{NASA}. Therefore, to avoid these  deteriorating effects, active debris removal (ADR) has become a challenging problem in the twenty-first century. Some indirect detection techniques of the  space debris have also been developed by different authors \cite{Kulikov, Sen,  Mukherjee, Acharya}.  Recently, a detailed investigation of solitons generated due to orbital space debris motion has been carried out by Truitt and Hartzell \cite{Truitt, TruittKP} through simulations using forced Korteweg-de Vries (KdV) equation and forced Kadomtsev-Petviashvili (KP) equation. They also discuss the feasibility of  observation of solitons caused by debris through current ground-based and in situ measurements.

The pioneering work of Sen et al \cite{Sen} proposes a new detection technique of charged
debris objects in the LEO plasma region through observation of precursor  solitons.  
In their subsequent works \cite{Sen1, SenMD, Sen2}, the numerical solutions for nonlinear ion acoustic waves  due to a moving charged source,  the molecular dynamic simulations for a charged object moving  in a
 strongly coupled dusty
 plasma to explore the existence of precursor solitonic pulses and dispersive shock waves has been discussed. Also,  the experimental observation of precursor solitons in a flowing dusty plasma are analyzed. The experimental observation of modifications in the propagation characteristics of precursor solitons caused by different
shapes and sizes of the object over which the dust fluid flows are reported in \cite{Sen3}. In our previous works \cite{Mukherjee, Acharya}  on space debris, pinned accelerated and curved solitary wave solutions due to space debris motion are discussed; providing a detailed understanding of the dynamical properties of solitary waves in the LEO region in $(1+1)$ and $(2+1)$ dimensions. In these works, we discuss an indirect detection technique for orbital debris objects by observation of changes in amplitudes and velocities of solitary waves. This detection technique can be compared with the earlier technique for space debris detection of Kulikov and Zak \cite{Kulikov} through observation of increase of amplitudes.  The effects of dust cloud on space debris dynamics, which lead to bending of dust ion acoustic solitary waves in the LEO region, are investigated in \cite{Acharya} in (2+1) dimension. The dynamical behaviour of nonlinear ion acoustic wave in  presence of sinusoidal
source debris term  in Thomas-Fermi plasmas  is  explored in \cite{TFMandi}.

In nearly all of the above explorations on space debris, effects of ambient magnetic field in the LEO region are neglected; except the work by Kumar and Sen \cite{Kumar}, who have observed both electrostatic and electromagnetic precursors as well as wakes due to a moving charge bunch in a plasma. In absence of this ambient magnetic field, the dynamics of ion acoustic solitary waves can represent appropriately the evolution of orbital space debris. In realistic plasma environment in the LEO region, that is being exposed to streaming solar wind plasma and, also, contains ionospheric plasma of the Earth \cite{Truitt}, the presence of magnetic field cannot be neglected. This is because this plasma region comes within the magnetosphere of the earth, and is being influenced by interplanetary magnetic fields as well. Therefore, in presence of this interpenetrating magnetic field in the LEO plasma region containing moving charged space debris particles, magnetohydrodynamic or hydromagnetic waves like magnetosonic waves can be more crucial than ion acoustic waves. This fact is further consolidated by the recent work of Kumar and Sen \cite{Kumar}; who have reported excitations of precursor magnetosonic solitons in ionospheric plasma condition due to debris. This work of Kumar and Sen is performed through particle in cell (PIC) simulations; along with the approximation of the ambient magnetic field in $Z$ direction only, which is not more realistic as desired. For dealing with these hydromagnetic waves, a magnetohydrodynamics (MHD) model will be useful rather than simple fluid models as done in \cite{Sen, Mukherjee, Acharya}. In particular, for explorations in space and astrophysical conditions involving magnetized plasmas, Hall MHD model is extremely useful; which takes into account the Hall current generated in the magnetized plasma behaving as a dielectric medium. When the characteristic length scales of a problem related to plasma
motion become shorter than or comparable to ion inertial lengths and time scales become shorter than or comparable to ion cyclotron periods, we have to use Hall magnetohydrodynamics instead of classical magnetohydrodynamics (MHD) \cite{Ruderman, Huba}. Hall plasma
is an anisotropic medium due to the presence of magnetic field, 
described by Hall MHD; that is used in various laboratory as well as astrophysical problems. It can be used to
describe magnetosonic waves in the solar atmosphere and in the magnetosphere of the Earth \cite{Ruderman}. Kadomtsev-Petviashvili (KP) equation was derived in \cite{Ruderman} to show the stability of slow and fast
magnetosonic solitons using Hall MHD. Recently, Bandyopadhyay et al. \cite{Bandyopadhyay} have analysed the in situ data collected from Magnetospheric Multiscale (MMS) space-craft in the context of Hall magnetohydrodynamics; revealing some interesting novel results with possible explanations in space plasma coditions. Therefore, taking into account these facts, the plasma in the LEO region can be dealt with Hall magnetohydrodynamics appropriately.  
Along with the solitary waves, the two dimensional  lump waves are also interesting localized structures in space plasma physics. The dynamics of magnetosonic lump waves in presence of charged debris seems to be an interesting problem; which has not been studied analytically till now as far as our knowledge goes. Since, the debris field may induce perturbations in the magnetosonic wave structures; hence they may be self-consistently related to each other like \cite{Acharya, Mukherjee}. In this work, we consider the dynamics of (2+1) dimensional localized magnetosonic waves in presence of charged space debris and indicate a possible indirect way of detection of debris objects.

 The article is organized in the following manner. The detailed derivation of the (2+1) dimensional nonlinear evolution equation for the magnetosonic wave in the form of forced KP equation is given in section- II. The dynamics of magnetosonic localized waves in terms of exact as well as approximate lump wave solutions in presence of debris field are discussed in section- III.  In section -IV, the main findings of this article are recapitulated along with discussions and possible applications. Conclusive remarks are provided in section -V  followed by acknowledgements and bibliography. 

\section{Derivation of (2+1) dimensional nonlinear evolution equation for the magnetosonic wave in presence of charged space debris} \label{NLEEDerivation}
We consider the propagation of nonlinear magnetosonic waves in the LEO plasma region in presence of the ambient magnetic field due to motion of orbital charged debris objects. The LEO region consists of a low temperature-low density plasma containing numerous space debris particles. In particular, this magnetized plasma can be categorized as Hall plasma like many space and astrophysical plasmas \cite{Bandyopadhyay}. The dynamics of these nonlinear waves can suitably be described by the equations of Hall magnetohydrodynamics (MHD) \cite{Ruderman}; which are given by
\begin{equation}
\frac{\partial \rho}{\partial t}+\vec{\nabla}.(\rho \vec{v})=0, \label{Continuity}
\end{equation}
\begin{equation}
\rho [\frac{\partial \vec{v}}{\partial t}+(\vec{v}.\vec{\nabla})\vec{v}]=-\beta \vec{\nabla}\rho+(\vec{\nabla} \times \vec{B}) \times \vec{B}-\vec{J_s} \times \vec{B}, \label{Momentum}
\end{equation}
\begin{equation}
\rho \frac{\partial \vec{B}}{\partial t}=\rho \vec{\nabla} \times (\vec{v} \times \vec{B})-\vec{\nabla} \times [(\vec{\nabla} \times \vec{B})\times \vec{B}], \label{Faraday}
\end{equation}
\begin{equation}
P=P_0{(\frac{\rho}{{\rho}_0})}^{\gamma}. \label{Pressure}
\end{equation}
The equations (\ref{Continuity}), (\ref{Momentum}), (\ref{Faraday}) and (\ref{Pressure}) represent the continuity equation, momentum equation, Ohm's law and adiabatic pressure law respectively. Here $\rho$ is mass density, $\vec{v}$ is the ($3D$) fluid velocity, $\vec{B}$ is the ambient magnetic field, ${\vec{J_s}}$ is the source current in the LEO plasma region arising due to the motion of charged space debris particles, P is the pressure in the plasma medium and $\gamma$ is the adiabatic constant. Similarly $\rho_0$ is the equilibrium mass density and $P_0$ is the equilibrium pressure. $\beta$ is a constant; which is given by
\begin{equation}
\beta=\frac{c^2_s}{v^2_A}, \label{beta}
\end{equation}
where $c_s$ is sound speed and $v_A$ is Alfven speed. When $c_s \leq v_A$, the value of $\beta$ varies from $0$ to $1$. On the other hand, when $c_s \geq v_A$, the value of $\beta$ exceeds $1$ starting from $1$ itself. As $\beta$ is the ratio of sound speed $c_s$ and Alfven speed $v_A$, its particular value for a given system can be determined by the ion density, ion temperature and strength of the ambient magnetic field in the system.
The normalizations used in the system of Hall MHD equations (\ref{Continuity}), (\ref{Momentum}), (\ref{Faraday}) and (\ref{Pressure}) are given by
\begin{equation}
\rho \longrightarrow \frac{\rho}{{\rho}_0};\, B \longrightarrow \frac{B}{B_0}, \label{HallNorm}
\end{equation}
where $B_0$ represents the equilibrium magnetic field present in the LEO plasma region. In terms of each component, the continuity equation (\ref{Continuity}) can be written as
\begin{equation}
\frac{\partial \rho}{\partial t}+\frac{\partial}{\partial x}(\rho v_x)+
\frac{\partial}{\partial y}(\rho v_y) +
\frac{\partial}{\partial z}(\rho v_z)=0. \label{Cont}
\end{equation}
Equation (\ref{Momentum}) can be decomposed into following three equations:
\begin{equation}
\rho[\frac{\partial v_x}{\partial t}+(v_x\frac{\partial}{\partial x}+ v_y\frac{\partial}{\partial y}+
v_z\frac{\partial}{\partial z})v_x]=-\beta \frac{\partial \rho}{\partial x}+[B_z(\frac{\partial B_x}{\partial z}-\frac{\partial B_z}{\partial x})-B_y(\frac{\partial B_y}{\partial x}-\frac{\partial B_x}{\partial y})]+J_{sz}B_y, \label{MomtX}
\end{equation}
\begin{equation}
\rho[\frac{\partial v_y}{\partial t}+(v_x\frac{\partial}{\partial x}+ v_y\frac{\partial}{\partial y}+
v_z\frac{\partial}{\partial z})v_y]=-\beta \frac{\partial \rho}{\partial y}+[B_x(\frac{\partial B_y}{\partial x}-\frac{\partial B_x}{\partial y})-B_z(\frac{\partial B_z}{\partial y}-\frac{\partial B_y}{\partial z})]-J_{sz}B_x, \label{MomtY}
\end{equation}
\begin{equation}
\rho[\frac{\partial v_z}{\partial t}+(v_x\frac{\partial}{\partial x}+ v_y\frac{\partial}{\partial y}+
v_z\frac{\partial}{\partial z})v_z]=-\beta \frac{\partial \rho}{\partial z}+[B_y(\frac{\partial B_z}{\partial y}-\frac{\partial B_y}{\partial z})-B_x(\frac{\partial B_x}{\partial z}-\frac{\partial B_z}{\partial x})]. \label{MomtZ}
\end{equation}
Similarly, equation (\ref{Faraday}) can be decomposed into following three equations:
$$
\rho \frac{\partial B_x}{\partial t}=\rho[\frac{\partial}{\partial y}(v_xB_y-v_yB_x)-\frac{\partial}{\partial z}(v_zB_x-v_xB_z)]-\lbrace \frac{\partial}{\partial y}[B_y(\frac{\partial B_z}{\partial y}-\frac{\partial B_y}{\partial z})-B_x(\frac{\partial B_x}{\partial z}-\frac{\partial B_z}{\partial x})]-$$ 
\begin{equation}
\frac{\partial}{\partial z}[B_x(\frac{\partial B_y}{\partial x}-\frac{\partial B_x}{\partial y})-B_z(\frac{\partial B_z}{\partial y}-\frac{\partial B_y}{\partial z})] \rbrace, \label{FaradX}
\end{equation}
$$
\rho \frac{\partial B_y}{\partial t}=\rho[\frac{\partial}{\partial z}(v_yB_z-v_zB_y)-\frac{\partial}{\partial x}(v_xB_y-v_yB_x)]-\lbrace \frac{\partial}{\partial z}[B_z(\frac{\partial B_x}{\partial z}-\frac{\partial B_z}{\partial x})-B_y(\frac{\partial B_y}{\partial x}-\frac{\partial B_x}{\partial y})]-$$ 
\begin{equation}
\frac{\partial}{\partial x}[B_y(\frac{\partial B_z}{\partial y}-\frac{\partial B_y}{\partial z})-B_x(\frac{\partial B_x}{\partial z}-\frac{\partial B_z}{\partial x})] \rbrace, \label{FaradY}
\end{equation}
$$
\rho \frac{\partial B_z}{\partial t}=\rho[\frac{\partial}{\partial x}(v_zB_x-v_xB_z)-\frac{\partial}{\partial y}(v_yB_z-v_zB_y)]-\lbrace \frac{\partial}{\partial x}[B_x(\frac{\partial B_y}{\partial x}-\frac{\partial B_x}{\partial y})-B_z(\frac{\partial B_z}{\partial y}-\frac{\partial B_y}{\partial z})]-$$ 
\begin{equation}
\frac{\partial}{\partial y}[B_z(\frac{\partial B_x}{\partial z}-\frac{\partial B_z}{\partial x})-B_y(\frac{\partial B_y}{\partial x}-\frac{\partial B_x}{\partial y})] \rbrace. \label{FaradZ}
\end{equation}
We assume that the nonlinear waves have weak transverse propagation. We consider that the source debris current resulting from the orbital charged debris particles and ambient magnetic field in the LEO plasma medium is approximated to be in $z$ direction. In order to derive the appropriate nonlinear evolution equation (NLEE) governing the dynamics of nonlinear waves using the well-known reductive perturbation technique (RPT) \cite{Sen}, the dependent variables are expanded as:
\begin{equation}
\rho=1+\epsilon {\rho}_1+{\epsilon}^2 {\rho}_2+O({\epsilon}^3),\label{RhoExpn}
\end{equation}
\begin{equation}
v_x=\epsilon {v_x}_1+{\epsilon}^2 {v_x}_2+O({\epsilon}^3),\label{VxExpn}
\end{equation}
\begin{equation}
v_y=\epsilon {v_y}_1+{\epsilon}^2 {v_y}_2+O({\epsilon}^3),\label{VyExpn}
\end{equation}
\begin{equation}
v_z={\epsilon}^{3/2} {v_z}_1+{\epsilon}^{5/2} {v_z}_2+O({\epsilon}^{7/2}),\label{VzExpn}
\end{equation}
\begin{equation}
B_x=cos (\alpha),
\label{BxExpn}
\end{equation}
\begin{equation}
B_y=sin (\alpha)+\epsilon {B_y}_1+{\epsilon}^2 {B_y}_2+O({\epsilon}^3),\label{ByExpn}
\end{equation}
\begin{equation}
B_z={\epsilon}^{3/2} {B_z}_1+{\epsilon}^{5/2} {B_z}_2+O({\epsilon}^{7/2}),\label{BzExpn}
\end{equation}
\begin{equation}
J_{sz}={\epsilon}^{5/2}J, \label{JExpn}
\end{equation}
\begin{equation}
P=P_0+\epsilon P_1+{\epsilon}^2P_2+O({\epsilon}^3), \label{PExpn} 
\end{equation}
where $\epsilon$ is a small dimensionless parameter characterizing the strength of nonlinearity in the system. Here $\alpha$ is the angle which the magnetic field $\vec{B}$ makes with $x$-axis.
For simplicity, it is assumed that the dependent variables do not change along $y$ direction. This yields $\vec{\nabla}=(\frac{\partial}{\partial x}, \frac{\partial}{\partial z})$, as $\frac{\partial}{\partial y}=0$. Since, we have considered weak transverse propagation, the scalings of $v_z, J_{sz}$ are weaker compared to other components.

 Other independent variables are rescaled as 
\begin{equation}
\xi={\epsilon}^{1/2}(x-at);\,\zeta=\epsilon z;\,\tau={\epsilon}^{3/2}t, a = {magnetosonic \ phase \ velocity},
\end{equation}
since we have considered weak propagation along transverse $z$ direction. After substituting these expanded and stretched variables in the contituity equation (\ref{Cont}), and equating the coefficients of different powers of $\epsilon$ to zero, we get:
\begin{equation}
O({\epsilon}^{3/2}):\, -a\frac{\partial {\rho}_1}{\partial \xi}+\frac{\partial {v_x}_1}{\partial \xi}=0, \label{ContTT}
\end{equation}
\begin{equation}
O({\epsilon}^{5/2}):\, -a\frac{\partial {\rho}_2}{\partial \xi}+\frac{\partial {\rho}_1}{\partial \tau}+\frac{\partial {v_x}_2}{\partial \xi}+\frac{\partial}{\partial \xi}({\rho}_1{v_x}_1)+\frac{\partial {v_z}_1}{\partial \zeta}=0. \label{ContFT}
\end{equation}
Then, we do the same procedure to  equations (\ref{MomtX}), (\ref{MomtY}) and (\ref{MomtZ}). From equation (\ref{MomtX}), we get:
\begin{equation}
O({\epsilon}^{3/2}):\, -a\frac{\partial {v_x}_1}{\partial \xi}=-\beta \frac{\partial {\rho}_1}{\partial \xi}-sin (\alpha) \frac{\partial {B_y}_1}{\partial \xi}, \label{MomtXTT}
\end{equation}
\begin{equation}
O({\epsilon}^{5/2}):\,-a{\rho}_1\frac{\partial {v_x}_1}{\partial \xi}-a\frac{\partial {v_x}_2}{\partial \xi}+\frac{\partial {v_x}_1}{\partial \tau}+{v_x}_1\frac{\partial {v_x}_1}{\partial \xi}=-\beta\frac{\partial {\rho}_2}{\partial \xi}-\beta \frac{\gamma -1}{2}\frac{\partial}{\partial \xi}({{\rho}_1}^2)-sin (\alpha) \frac{\partial {B_y}_2}{\partial \xi}-{B_y}_1\frac{\partial {B_y}_1}{\partial \xi}+J sin (\alpha). \label{MomtXFT} 
\end{equation}
In the same way, equation (\ref{MomtY}) yields:
\begin{equation}
O({\epsilon}^{3/2}):\,-a\frac{\partial {v_y}_1}{\partial \xi}=cos (\alpha) \frac{\partial {B_y}_1}{\partial \xi},\label{MomtYTT}
\end{equation}
\begin{equation}
O({\epsilon}^{5/2}):\,-a\frac{\partial {v_y}_2}{\partial \xi}-a{\rho}_1\frac{\partial {v_y}_1}{\partial \xi}+\frac{\partial {v_y}_1}{\partial \tau}+{v_x}_1\frac{\partial {v_y}_1}{\partial \xi}= cos (\alpha) \frac{\partial {B_y}_2}{\partial \xi}-J cos (\alpha). \label{MomtYFT}
\end{equation}
Equation (\ref{MomtZ}) gives:
\begin{equation}
O({\epsilon}^2):\,-a\frac{\partial {v_z}_1}{\partial \xi}=-\beta \frac{\partial {\rho}_1}{\partial \zeta}-sin (\alpha) \frac{\partial {B_y}_1}{\partial \zeta}+cos (\alpha) \frac{\partial {B_z}_1}{\partial \xi}. \label{MomtZT}
\end{equation}
Equation (\ref{FaradX}) yields:
\begin{equation}
O({\epsilon}^{5/2}):\,-cos (\alpha) \frac{\partial {v_z}_1}{\partial \zeta}+cos (\alpha) \frac{{\partial}^2 {B_y}_1}{\partial \zeta \partial \xi} = 0. \label{FaradXFT}
\end{equation}
Equation (\ref{FaradY}) yields:
\begin{equation}
O({\epsilon}^{3/2}):\,-a\frac{\partial {B_y}_1}{\partial \xi}=-sin (\alpha) \frac{\partial {v_x}_1}{\partial \xi}+cos (\alpha) \frac{\partial {v_y}_1}{\partial \xi},\label{FaradYTT}
\end{equation}
$$
O({\epsilon}^{5/2}):\,-a\frac{\partial {B_y}_2}{\partial \xi}-a {\rho}_1 \frac{\partial {B_y}_1}{\partial \xi}+\frac{\partial {B_y}_1}{\partial \tau}=-sin (\alpha) \frac{{v_z}_1}{\partial \zeta}-\frac{\partial}{\partial \xi}({v_x}_1{B_y}_1)-sin (\alpha) \frac{{v_x}_2}{\partial \xi}+cos (\alpha) \frac{\partial {v_y}_2}{\partial \xi}-sin (\alpha) {\rho}_1\frac{\partial {v_x}_1}{\partial \xi}+$$
\begin{equation}
cos (\alpha) {\rho}_1 \frac{\partial {v_y}_1}{\partial \xi}+cos (\alpha) \frac{{\partial}^2{B_z}_1}{\partial {\xi}^2}. \label{FaradYFT}
\end{equation}
Finally, Eq. (\ref{FaradZ}) yields:
\begin{equation}
O({\epsilon}^2):\,-a\frac{\partial {B_z}_1}{\partial \xi}=cos (\alpha) \frac{\partial {v_z}_1}{\partial \xi}-cos (\alpha) \frac{{\partial}^2{B_y}_1}{\partial {\xi}^2}. \label{FaradZT}
\end{equation}
From equations (\ref{ContTT}), (\ref{MomtXTT}), (\ref{MomtYTT}) and (\ref{FaradYTT}), we get after some simplifications:
\begin{equation}
{v_x}_1=a{\rho}_1;\,{v_y}_1=-\frac{a \,sin (\alpha) cos (\alpha)}{a^2-cos^2 (\alpha)}{\rho}_1;\,{B_y}_1=\frac{a^2 sin (\alpha)}{a^2-cos^2 (\alpha)}{\rho}_1, \label{Vx1Vy1By1}
\end{equation}
along with the dispersion relation of the magnetosonic waves:
\begin{equation}
(a^2-\frac{{c_s}^2}{{v_A}^2})[a^2-cos^2 (\alpha)]=a^2 sin^2 (\alpha).\label{DispRel}
\end{equation}
From the above dispersion relation (\ref{DispRel}), the phase velocity $a$ is easily obtained as:
\begin{equation}
a^2=\frac{(1+\beta) \pm \sqrt{{(1+\beta)}^2-4\beta cos^2 (\alpha)}}{2}. \label{DispRel1}
\end{equation}
The plus and minus signs in the above expression for $a$ correspond to fast and slow magnetosonic waves respectively. The variation of $a$ against angle of propagation $\alpha$ for a fixed value of $\beta$ are shown in Figure-1 for both slow and fast magnetosonic waves.
%\begin{figure}[hbt!]
%\centering
%\includegraphics[width=10cm]{Slowa.png}
%\vspace{1cm}
%\caption{Variation of phase velocity $a$ against $\alpha$ for $\beta =0.8$ in case of slow magnetosonic waves}
%\end{figure}
\begin{figure}[hbt!]
\centering
\includegraphics[width=10cm]{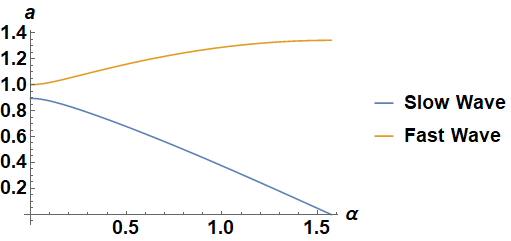}
\vspace{1cm}
\caption{Variation of phase velocity $a$ against $\alpha$ for $\beta =0.8$ in case of both slow and fast magnetosonic waves}
\end{figure}
 Now we separate the first and second order variables from equations (\ref{ContFT}), (\ref{MomtXFT}), (\ref{MomtYFT}) and (\ref{FaradYFT}) to obtain after some simplifications
\begin{equation}
a\frac{\partial {\rho}_2}{\partial \xi}-\frac{\partial {v_x}_2}{\partial \xi}=\frac{\partial {\rho}_1}{\partial \tau}+2a{\rho}_1\frac{\partial {\rho}_1}{\partial \xi}+\frac{\partial {v_z}_1}{\partial \zeta}, \label{AFst}
\end{equation}
\begin{equation}
\beta \frac{\partial {\rho}_2}{\partial \xi}-a\frac{\partial {v_x}_2}{\partial \xi}+sin (\alpha) \frac{\partial {B_y}_2}{\partial \xi}=-a\frac{\partial {\rho}_1}{\partial \tau}-\frac{{(a^2-\beta)}^2}{sin^2 (\alpha)}{\rho}_1\frac{\partial {\rho}_1}{\partial \xi}-\beta(\gamma -1){\rho}_1\frac{\partial {\rho}_1}{\partial \xi}+J sin (\alpha), \label{BFst}
\end{equation}
\begin{equation}
a\frac{\partial {v_y}_2}{\partial \xi}+cos (\alpha) \frac{\partial {B_y}_2}{\partial \xi}=-\frac{(a^2-\beta)cos (\alpha)}{a sin (\alpha)}\frac{\partial {\rho}_1}{\partial \tau}-Jcos (\alpha), \label{CFirst}
\end{equation}
\begin{equation}
a \frac{\partial {B_y}_2}{\partial \xi}-sin (\alpha) \frac{\partial {v_x}_2}{\partial \xi}+cos (\alpha) \frac{\partial {v_y}_2}{\partial \xi}=\frac{a^2-\beta}{sin (\alpha)}\frac{\partial {\rho}_1}{\partial \tau}+\frac{a^4-a^2\beta+a^2-\beta cos^2 (\alpha)}{a sin (\alpha)}{\rho}_1\frac{\partial {\rho}_1}{\partial \xi}+sin (\alpha) \frac{\partial {v_z}_1}{\partial \zeta}-cos (\alpha) \frac{{\partial}^2{B_z}_1}{\partial {\xi}^2}, \label{DFirst}
\end{equation}
where we have used the relations given by equation (\ref{Vx1Vy1By1}). Now, for simplicity of further calculations, we use the following notations for the LHSs of the equations (\ref{AFst}), (\ref{BFst}), (\ref{CFirst}) and (\ref{DFirst}) containing only second order variables: 
\begin{equation}
P=a\frac{\partial {\rho}_2}{\partial \xi}-\frac{\partial {v_x}_2}{\partial \xi};\,Q=\beta \frac{\partial {\rho}_2}{\partial \xi}-a\frac{\partial {v_x}_2}{\partial \xi}+sin (\alpha) \frac{\partial {B_y}_2}{\partial \xi}, \label{AB}
\end{equation}
\begin{equation}
R=a\frac{\partial {v_y}_2}{\partial \xi}+cos (\alpha) \frac{\partial {B_y}_2}{\partial \xi};\,S=a \frac{\partial {B_y}_2}{\partial \xi}-sin (\alpha) \frac{\partial {v_x}_2}{\partial \xi}+cos (\alpha) \frac{\partial {v_y}_2}{\partial \xi}. \label{CD}
\end{equation}
Therefore, the variables $P,\,Q,\,R$ and $S$ defined by the above equations (\ref{AB}) and (\ref{CD}) are functions of derivatives of only second order variables as specified in these equations. The above equations (\ref{AB}) and (\ref{CD}) can be simplified, after using the dispersion relation (\ref{DispRel}), to give a relation :
\begin{equation}
(a^2-1)P-\frac{a^2-cos^2 (\alpha)}{a}Q-\frac{sin (\alpha) cos (\alpha)}{a}R+sin (\alpha) S=0. \label{ABCDEqn}
\end{equation}
Differentiating the above equation (\ref{ABCDEqn}) partially with respect to $\xi$ once, and substituting equations (\ref{AFst}), (\ref{BFst}), (\ref{CFirst}) and (\ref{DFirst}), we obtain after some rigorous calculations using equations (\ref{MomtZT}) and (\ref{FaradZT}):
%(\ref{FaradXFT}) 
$$\frac{\partial}{\partial \xi} \lbrace \frac{\partial {\rho}_1}{\partial \tau}+\frac{a(a^2-\beta)}{2(2a^2-1-\beta)}[3+(\gamma+1)\frac{a^2-1}{a^2-\beta}]{\rho}_1 \frac{\partial {\rho}_1}{\partial \xi}+\frac{a(a^2-\beta)cos^2 (\alpha)}{2[cos^2 (\alpha)-a^2](2a^2-1-\beta)}\frac{{\partial}^3{\rho}_1}{\partial {\xi}^3}\rbrace-$$
\begin{equation}
\frac{a^3}{2(1+\beta-2a^2)}\frac{{\partial}^2{\rho}_1}{\partial {\zeta}^2}=\frac{a^2-2cos^2 (\alpha)}{a}sin (\alpha) \frac{\partial J}{\partial \xi}, \label{FinalKP}
\end{equation}
which is the final nonlinear evolution equation in the form of forced Kadomtsev-Petviashvili (fKP) equation. In compact form, this fKP equation can be written as
\begin{equation}
\frac{\partial}{\partial \xi}(\frac{\partial {\rho}_1}{\partial \tau}+N {\rho}_1\frac{\partial {\rho}_1}{\partial \xi}+D \frac{{\partial}^3{\rho}_1}{\partial {\xi}^3})- T \frac{{\partial}^2{\rho}_1}{\partial {\zeta}^2}=E \frac{\partial J}{\partial \xi}, \label{FinalKP1}
\end{equation}
where $N, \,D, \,T$ and $E$ denote the nonlinear, dispersive, $2D$ and external forcing coefficients in the forced KP equation (\ref{FinalKP}) respectively, and are given by
$$N=\frac{a(a^2-\beta)}{2(2a^2-1-\beta)}[3+(\gamma+1)\frac{a^2-1}{a^2-\beta}];\,D=\frac{a(a^2-\beta)cos^2 (\alpha)}{2[cos^2 (\alpha)-a^2](2a^2-1-\beta)};$$ 
\begin{equation}
T=\frac{a^3}{2(1+\beta-2a^2)};\,E=\frac{a^2-2cos^2 (\alpha)}{a}sin (\alpha).
\end{equation} 
Depending on the sign of coefficients in (\ref{FinalKP1}), KP equation can be of two types as KP-I and KP-II having different properties of solutions. In the following sections, we will analyze its different interesting solutions to explain the wave dynamics.

\section{Dynamics of  magnetosonic lump waves in presence of charged space debris} 
At  first sight, the nonlinear evolution equation (\ref{FinalKP1}) looks like a forced KPI equation \cite{Yong}. For a detailed analysis of the nature of the forced KP equation (\ref{FinalKP1}), we plot the coefficients $N,\,D$, $T$ and $E$ against the angle of propagation $\alpha$ for constant $\beta$. This is as shown in the Figure-2, 3.  From the figure, it is clear that $N,\,D$ and $T$ are all positive whereas $E$ is negative for slow magnetosonic waves. On the other hand, for fast magnetosonic waves, only $N$ is positive whereas $D$ and $T$ are negative, and $E$ is negative for lower values of $\alpha$ whereas it turns positive for higher values of $\alpha$.

Therefore, for the case of slow magnetosonic waves, the nonlinear evolution equation (\ref{FinalKP1}) becomes the forced KPI equation as $T$ is positive.  In order to determine the stability behaviour of both slow and fast magnetosonic solitary waves, we follow the recent work done by Ruderman \cite{Ruderman} on stability of magnetosonic solitons in Hall plasmas. From his analysis of stability of magnetosonic solitons, it is concluded that magnetosonic solitons are stable with respect to transverse perturbations provided they are governed by a KPII equation, and they are unstable with respect to transverse perturbations provided they are governed by a KPI equation. For the case of KPI equation, rational algebraic soliton or lump wave becomes stable; which is explored in detail in the subsequent sections. 
From the above stability analysis of Ruderman, it is obvious that the slow magnetosonic solitary waves are unstable with respect to transverse perturbations as their dynamical evolution is governed by a forced KPI equation.  For the fast magnetosonic wave, the evolution equation can be transformed to forced KP-I equation by a simple transformation; i.e, $\tau \rightarrow -\tau;\, \rho_1 \rightarrow -\rho_1$. Hence for both slow and fast magnetosonic waves, the localized rational lump solutions are stable.

In recent years, Kumar and Sen \cite{Kumar} have observed excitations of magnetosonic waves in presence of charged space debris in the ionospheric plasma after approximating the ambient magnetic field to be in $Z$ direction. Most interestingly, the curving nature of magnetosonic waves is reported in their work when a $2D$ circular source is considered. Their work is performed through particle in cell (PIC) simulation. No detailed analytical explanation of this spectacular phenomenon has been reported till now. In our recent work on space debris \cite{Acharya}, the bending phenomenon of electrostatic solitary waves is explained analytically; where we have derived a special exact curved  solitary wave solution of the nonlinear evolution equation. We have not considered the ambient magnetic field in the LEO region in this  work. Therefore, in order to explain the curving nature of magnetosonic  waves as reported by Kumar and Sen \cite{Kumar}, we need to derive a special exact curved solitary wave solution that we plan to do in future.

\begin{figure}[hbt!]
\centering
\includegraphics[width=13cm]{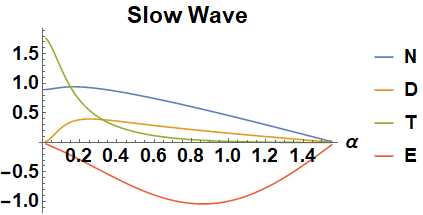}

\vspace{1cm}

\caption{Variation of the nonlinear, dispersive, $2D$ and forcing coefficients $N$, $D$, $T$ and $E$ respectively in forced KP equation (\ref{FinalKP1}) against angle of propagation $\alpha$ for the parameters $\beta=0.8$ and $\gamma =1$. In this figure, $\alpha$ is in radian unit. This figure is associated with slow magnetosonic waves }

\end{figure}

\begin{figure}[hbt!]
\centering
\includegraphics[width=13cm]{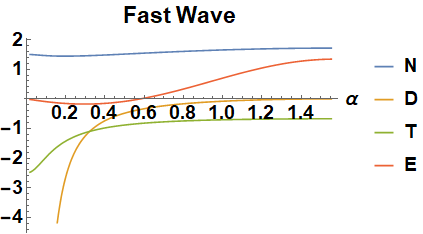}

\vspace{1cm}

\caption{Variation of the nonlinear, dispersive, $2D$ and forcing coefficients $N$, $D$, $T$ and $E$ respectively in forced KP equation (\ref{FinalKP1}) against angle of propagation $\alpha$ for the parameters $\beta=0.8$ and $\gamma =1$. In this figure, $\alpha$ is in radian unit. This figure is associated with fast magnetosonic waves }

\end{figure}

Now, we discuss some exact as well as approximate lump wave solutions of the nonlinear evolution equation (\ref{FinalKP1}) in presence of forcing debris term. As discussed before, the magnetosonic lump waves are stable with respect to transverse perturbations  for both fast and slow magnetosonic waves.  

Hence for slow magnetosonic waves, the evolution equation is given by forced KP-I equation given as :
\begin{equation}
\frac{\partial}{\partial \xi}(\frac{\partial {\rho}_1}{\partial \tau}+N {\rho}_1\frac{\partial {\rho}_1}{\partial \xi}+D \frac{{\partial}^3{\rho}_1}{\partial {\xi}^3})- T \frac{{\partial}^2{\rho}_1}{\partial {\zeta}^2}= -|E| \frac{\partial J}{\partial \xi}, \label{FinalKP1r}
\end{equation}
where $N, \, D$ and $T$ are positive constants, and $|E|$ denotes the magnitude of E. In order to obtain a more  convenient form of the nonlinear evolution equation (\ref{FinalKP1r}), we use the following redefinition of variables:
\begin{equation}
{\rho}_1={(\frac{6D}{N})}U;\,\xi = X;\,\zeta=\sqrt{\frac{T}{3D}} Z;\,\tau= (1/D) T;\,J=- \frac{D A}{|E|} F. \label{FrameTrans}
\end{equation}
Then, the nonlinear evolution equation takes the form:
\begin{equation}
\frac{\partial}{\partial X}(\frac{\partial U}{\partial T}+6U\frac{\partial U}{\partial X}+\frac{{\partial}^3U}{\partial X^3})-3\frac{{\partial}^2U}{\partial Z^2}=\frac{\partial F}{\partial X}, \label{FinalKPTrans}
\end{equation}
which is nothing but the forced KPI equation \cite{Yong}. In the similar fashion, we can derive such forced KP-I equation for fast magnetosonic wave with some transformation of variables as discussed before. In the following calculations, we analyze this equation (\ref{FinalKPTrans}) to explore its various solutions.
\subsection{Exact lump wave solution with a definite self-consistent forcing function}
Before discussing the exact accelerated lump wave solutions which are the main findings of our work, we will discuss a special exact lump solution of (\ref{FinalKPTrans}), where forcing term satisfies a constraint equation.
 In the last decade, Yong et al. \cite{Yong} have reported a self-consistent model for forced KPI equation, namely KPIESCS, i.e. KPI equation with a self-consistent source, where the source or forcing term obeys a specific constraint condition. Following their work, the self-consistent forcing function, in our work, is taken to be of the form:
\begin{equation}
F=-8\frac{\partial}{\partial X}(\Omega {\Omega}^*), \label{omega_for}
\end{equation}
where $\Omega$ denotes a complex variable, and is a function of $X,\,Z$ and $T$, i.e. $\Omega=\Omega(X,Z,T)$. Then, our nonlinear evolution equation (\ref{FinalKPTrans}) becomes
\begin{equation}
{[U_T+6UU_X+U_{XXX}+8{(\Omega {\Omega}^*)}_X]}_X-3U_{ZZ}=0. \label{Scaled_KP1}
\end{equation}
In order to solve the above equation (\ref{Scaled_KP1}), Yong et al. have taken a constraint condition to be satisfied by forcing function $F$ and the wave $U$ as
\begin{equation}
i{\Omega}_Z={\Omega}_{XX}+U\Omega. \label{Cons_lump}
\end{equation}
For implementation of Hirota bilinear method as done by Yong et al., we proceed further by applying a transformation $U=2{(ln H)}_{XX}$ and $\Omega = \frac{G}{H}$ to equations (\ref{Scaled_KP1}), and (\ref{Cons_lump}) to obtain the bilinear equation
\begin{equation}
(D_XD_T+{D_X}^4-3{D_Z}^2)H.H+8(GG^*-H^2)=0, \label{D1}
\end{equation}
 \begin{equation}
 (iD_Z-{D_X}^2)G.H=0, \label{D2}
 \end{equation}
 where $D$ is the famous Hirota bilinear operator. In order to investigate lump solutions for the above bilinear equations (\ref{D1}), and (\ref{D2}), we assume 
 \begin{eqnarray}
 H=1+{{\xi}_1}^2+{{\xi}_2}^2,\label{H_assmp}\\
 G=G_R+iG_I,  \label{G_assmp}
 \end{eqnarray}
following \cite{Yong}, where ${\xi}_1$ and ${\xi}_2$ are two new variables, and $G_R$ and $G_I$ are the real and imaginary parts of the complex variable $G$ respectively. Here $'i'$ denotes the imaginary  number. Our assumption for $H$ guarantees analyticity and rational localization of solutions. These new variables are defined in terms of the old variables as
 \begin{eqnarray}
 {\xi}_1=a_1X+a_2Z+a_3T+a_4;\, {\xi}_2=a_5X+a_6Z+a_7T+a_8, \label{xi_1_2}\\
G_R=b_0+b_1{\xi}_1+b_2{\xi}_2+b_3{{\xi}_1}^2+b_4{{\xi}_2}^2;\,G_I=c_0+c_1{\xi}_1+c_2{\xi}_2+c_3{{\xi}_1}^2+c_4{{\xi}_2}^2, \label{G_RI} 
 \end{eqnarray}
 where $a_i\,(1\leq i \leq 8)$, $b_j\,(0 \leq j \leq 4)$, and $c_l\,(0 \leq l \leq 4)$ are real constants to be determined subsequently. According to Yong et al., these real constants are related by the following expressions:
$$a_3=\frac{({a_2}^2-{a_1}^4)[3{({a_1}^4+{a_2}^2)}^2+16{a_1}^4]}{a_1{({a_1}^4+{a_2}^2)}^2},\,a_5=0,\,a_6={a_1}^2,\,a_7=\frac{2a_1a_2[3{({a_1}^4+{a_2}^2)}^2-16{a_1}^4]}{{({a_1}^4+{a_2}^2)}^2},$$

$$\,b_0=\frac{b_3{({a_2}^2-3{a_1}^4)}}{{a_1}^4+{a_2}^2},\,b_1=kc_1,\,b_2=kc_2,\,b_4=b_3,$$
\begin{equation}
\,c_0=-kb_0,\,c_1=-\frac{4{a_1}^2a_2b_3}{{a_1}^4+{a_2}^2},\,c_2=\frac{4{a_1}^4b_3}{{a_1}^4+{a_2}^2},\,c_3=-kb_3, \label{Par_cons}
\end{equation}
where $k$ is a constant which requires to satisfy ${b_3}^2(1+k^2)=1$. The localization of the associated solutions is also guaranteed here because the non-zero determinant condition as illustrated in \cite{Ma} is satisfied, i.e. $a_1a_6-a_2a_5 \neq 0$.

In explicit form, the solution $U$ is obtained in original variables $X,\,Z$ and $T$ as:
%Following the values of constants used in \cite{Yong}, one admissible lump solution $U$ typically looks like
\begin{equation}
U(X,Z,T)=4\frac{-{(a_1X+a_2Z+a_3T+a_4)}^2+{(a^2_1Z+a_7T+a_8)}^2+1}{{[{(a_1X+a_2Z+a_3T+a_4)}^2+{(a^2_1Z+a_7T+a_8)}^2+1]}^2}, \label{Lump_c}
\end{equation}
where we have used equation (\ref{Par_cons}) for substitution of $a_5=0$ and $a_6=a^2_1$ in writing the above solution (\ref{Lump_c}). In particular, there are multiple sets of values of parameters satisfying the condition (\ref{Par_cons}); some of which are as given in \cite{Yong}. Putting these admissible values of parameters in the solution (\ref{Lump_c}), the final lump wave solution can be obtained as discussed by Yong et al. \cite{Yong} in detail. 
%The above solution (\ref{Lump_c}) can be pictured from Figure-4 at time $T=0$ on $X-Z$ plane.
 More explicitly, the above solution (\ref{Lump_c}) moves with a constant velocity; which is given by
\begin{equation}
\vec{V}=(V_X,V_Z);\,V_X=\frac{a_2a_7-a_3a^2_1}{a^3_1} \, \ and\ \,V_Z=-\frac{a_7}{a^2_1}\, \label{YongVel}
\end{equation}
where $V_X$ and $V_Z$ denote $X$ and $Z$ components of velocity $\vec{V}$ of lump wave solution (\ref{Lump_c}) respectively. Similarly the maximum amplitude of this lump wave solution at the  (\ref{Lump_c}) is given by
\begin{equation}
A_{max}=4;\,A_{org}=4\frac{-{(a_3T+a_4)}^2+{(a_7T+a_8)}^2+1}{{[{(a_3T+a_4)}^2+{(a_7T+a_8)}^2+1]}^2}, \label{YongAmp}
\end{equation} 
where $A_{max}$ denote the maximum amplitude of the solution (\ref{Lump_c}), and $A_{org}$ represent the amplitude of the solution (\ref{Lump_c}) at origin. Thus, the time-dependence of the amplitude can be visualized from $A_{org}$.
%\begin{figure}[hbt!]
%\centering
%\includegraphics[width=8cm]{Yong.png}
%\caption{Lump wave solution as represented by equation (\ref{Lump_c}) on $X-Z$ plane for $T=0$}
%\end{figure}
% and, then, plot it at origin, i.e. at $X=0$ and $Z=0$ against time $T$; which is as shown in figure \ref{Lump_soliton2}.
% The complete dynamical evolution of the lump solution for our nonlinear evolution equation \ref{FinalKPTrans} can be picturized by the superposition of the dynamical behaviours as represented in figures \ref{Lump_soliton1}, and \ref{Lump_soliton2}.
 \subsection{Exact pinned accelerated lump wave solution}
 As reported in \cite{Sen}, the debris field may induce perturbations in the nonlinear waves. Hence they may be self consistently related to each other as discussed in \cite{Mukherjee, Acharya}. In this work, we consider such kind of dependence where $F = F(U)$, which lead to pinned wave solutions as discussed in \cite{Sen, Acharya}. Since, the magnetosonic wave and the debris wave move together with same velocity, they are thought to be ``pinned'' together. For simplicity if we consider $F = a \ U_X,$
 where $a$ is a constant, we would get an exact pinned lump wave solution that moves  with a constant velocity. 
 Now, similar to the previous case, if  we assume that the amplitude of the
 forcing function $F$ is time dependent and it is related self-consistently with  $U$ as
\begin{equation}
F=A(T)\frac{\partial U}{\partial X}, \label{fURel}
\end{equation}
where $A(T)$ is a function of time, we would get an accelerated pinned lump solution.
On moving to a new frame of reference defined by
\begin{equation}
\tilde{X}=X+B(T);\,\tilde{Z}=Z;\,\tilde{T}=T, \label{FrameTrans1}
\end{equation}
we get after some simplifications from (\ref{FinalKPTrans}):
\begin{equation}
\frac{\partial}{\partial \tilde{X}}(\frac{\partial U}{\partial \tilde{T}}+6U\frac{\partial U}{\partial \tilde{X}}+\frac{{\partial}^3U}{\partial {\tilde{X}}^3})-3\frac{{\partial}^2U}{\partial {\tilde{Z}}^2}=0, \label{FinalKPTrans1}
\end{equation}
where we have chosen $B(T)=\int A(T)dT$. Therefore, finally we got an unforced KPI equation (\ref{FinalKPTrans1}) as shown above. It is known that this unforced KPI equation (\ref{FinalKPTrans1}) admits two dimensional rational algebraic lump wave solution, which is given by
\begin{equation}
U=4\frac{-{[\tilde{X}+m\tilde{Z}+3(m^2-n^2)\tilde{T}]}^2+n^2{(\tilde{Z}+6m\tilde{T})}^2+\frac{1}{n^2}}{{\lbrace {[\tilde{X}+m\tilde{Z}+3(m^2-n^2)\tilde{T}]}^2+n^2{(\tilde{Z}+6m\tilde{T})}^2+\frac{1}{n^2} \rbrace}^2}. \label{PinnedLump}
\end{equation}
The above equation (\ref{PinnedLump}) represents the lump solution in new variables. Transferring this into the original variables, i.e. $X,\,Z$ and $T$, one can get the acceleration of the lumps due to presence of the term $\int A(T) dT$ in the arguments. This solution looks like
\begin{equation}
U=4\frac{-{[X+\int A(T)dT+mZ+3(m^2-n^2)T]}^2+n^2{(Z+6mT)}^2+\frac{1}{n^2}}{{\lbrace {[X+\int A(T)dT+mZ+3(m^2-n^2)T]}^2+n^2{(Z+6mT)}^2+\frac{1}{n^2} \rbrace}^2}, \label{PinnedLump1}
\end{equation}
which represents the final form of the accelerated lump solution. In absence of $F$, the lump solution will decay exponentially to both spatial directions as seen in Figure-4. But in presence of F of the form (\ref{fURel}), such lump solutions will be accelerated that can be seen in  Figure-5, where $U$ is plotted in $X-T$ plane.  
\begin{figure}[hbt!]
\centering
\includegraphics[width=13cm]{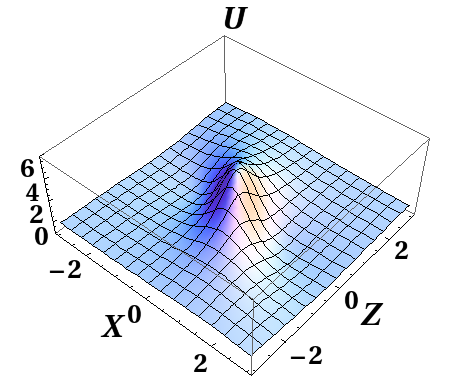}

\caption{ The 3D plot of the lump solution U (\ref{PinnedLump1}) without the debris term ($F=0$) in $X-Z$ plane for time $T=0,$ for the choice of $m=0.5,n=1.$ It can be seen that the lump solution decays in all directions.} 
\end{figure}
\begin{figure}

\includegraphics[width=13cm]{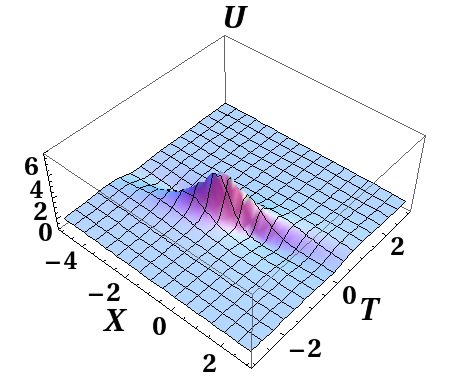}
 
%\vspace{1cm}

 (a) Magnetosonic lump solution $U$ (\ref{PinnedLump1}) in $X-T$ plane at $Z=1,$ for $F=0$ and $n=1,m=0.5$.

%\vspace{1cm}

\includegraphics[width=13cm]{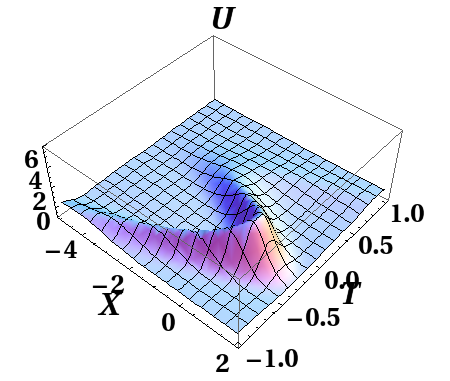}

%\vspace{1cm}

(b) Acceleration of the magnetosonic lump solution $U$ (\ref{PinnedLump1}) in $X-T$ plane at $Z=1,$ for $A(T)=8\cos{4T}$ and $n=1,m=0.5$.
%\noindent FIG.3:
\caption{ The figures show the 3D plots of lump solution $U$ (\ref{PinnedLump1}) on $X-T$ plane without and with forcing $F$, for $Z=1$.
 Due to the presence of nonlinear function $A(T)$ in the solution (\ref{PinnedLump1}), acceleration of the lump occurs. This can be seen from the distortion in the lump in $X-T$ plane in subfigure (b).}
\end{figure}

The velocity of the magnetosonic  and the debris lump solutions can be calculated analytically as follows.

If we define two variables as 
\begin{equation}
 \psi_1 = [{X}+ \int A(T)dT + m {Z}+3(m^2-n^2) {T} ], \ \psi_2 = n{[{Z}+6m {T}]}, 
\end{equation}
then the solution (\ref{PinnedLump1}) can be written as 
\begin{equation}
 U = 4 \frac{[- \psi_1^2 + \psi_2^2 + 1/n^2]}{[\psi_1^2 + \psi_2^2 + 1/n^2]^2}.
\end{equation}
The maximum amplitude of the lump solution $U$ is attained at $\psi_1 = \psi_2 =0.$ Hence, the velocity of the  
maximum amplitude point of the wave can be determined as  
\begin{equation}
 \frac{d \psi_2}{dT} = 0 \longrightarrow V_{UZ} = \frac{dZ}{dT} = -6m.
\end{equation}
Similarly, we can get 
\begin{equation}
 \frac{dX}{dT} = V_{UX}=3(m^2+n^2)-A(T),
\end{equation}
where,
$
\vec{V}_U=(V_{UX},V_{UZ})$
 denote the velocity with which the accelerated lump wave solution (\ref{PinnedLump1}) moves, and $V_{UX}$ and $V_{UZ}$ represent the $X$ and $Z$ components of $\vec{V}_U$ respectively. Hence, the acceleration associated with this solution  (\ref{PinnedLump1}) can be evaluated as:
\begin{equation}
\vec{W}_U=(W_{UX},W_{UZ});\,W_{UX}=-\frac{dA(T)}{dT};\,W_{UZ}=0, \label{AccUaVel}
\end{equation}
where $\vec{W}_U$ denote the acceleration of the solution (\ref{PinnedLump1}), and $W_{UX}$ and $W_{UZ}$ represent its $X$ and $Z$ components respectively. From the above equation (\ref{AccUaVel}), it is clear that acceleration of lump wave $U$ is coming due to presence of the term $A(T)$  in equation (\ref{PinnedLump1}); which can be understood from the distortion of the lump solution on X-T plane as seen in Figure-5(b).
The maximum amplitude of the accelerated lump wave solution (\ref{PinnedLump1}) is attained at $\psi_1=0=\psi_2$ and given as = $4 n^2.$ The maximum amplitudes of $U$ and $F$ are very different from each other, so that that can be detected separately using modern techniques. 
% For $m=0,$ the maximum amplitude will look like
% \begin{equation}
% U_{amp}=4\frac{-{[\int A(T)dT-3n^2T]}^2 +\frac{1}{n^2}}{{\lbrace {[\int A(T)dT-3n^2T]}^2+\frac{1}{n^2} \rbrace}^2}, \label{Uamp}
% \end{equation}
% which is time dependent.

The forcing function (\ref{fURel}) can  explicitly be written as:
\begin{equation}
F=-8A(T)\frac{{\psi}_1 (-{\psi}^2_1+3{\psi}^2_2+\frac{3}{n^2})}{{({\psi}^2_1+{\psi}^2_2+\frac{1}{n^2})}^3}, \label{FExpanded}
\end{equation}
where we have used the exact accelerated lump wave solution (\ref{PinnedLump1}) for substitution of $U$ in the expression of forcing function (\ref{fURel}).
It can be noted that  at $\psi_1=\psi_2 =0$ the forcing function $F$ becomes zero, whereas $U$ is maximum ($=4n^2$). In a similar way, the velocity and acceleration of the null point (where $F=0$) of the forcing function can
be calculated. It turns out that the velocity and acceleration of the point $\psi_1=\psi_2 = 0$ of the forcing function $F$, is same as that of $U$. Hence they can be called as pinned accelerated lump wave solutions because the propagate together.

The entire dynamics of $U$ and $F$ can be seen from Figure-6. If we take $m=0$ for any spacefic value of n, the velocity along $Z$ axis vanishes. Hence both the lump waves $U$ and $F$ move along $X$ axis.  Now if we concentrate on the point $P$ : $\psi_1 =0, \psi_2 = 0$ of both $F$ and $U$ and see their dynamics at different times ($T=0$ and $T=1$), we can observe that they move simultaneously as seen in Figure-6. It should also be noted here that forcing debris function is originally derived to be of Gaussian nature by Truitt et al. \cite{Truitt, TruittKP} in both one-dimension and two-dimensions. We know that Gaussian function in two-dimensions behaves like a lump soliton; which decays in all spatial directions. Therefore, from the plots of forcing debris function $F$ in Figure-6, it is clear that our assumption of considering $F$ to be of the form given in equation (\ref{fURel}) is desirably consistent for space debris particles. This is because this forcing debris function $F$, as visualized from Figure-6, behaves like a two-dimensional Gaussian function that is similar to a lump soliton as stated previously.

 In order to explore more physical properties of space debris particles, which are represented by the forcing function $F$, we further analyze the expression (\ref{FExpanded}) for $F$. For this, we first proceed to find its extremum points through partial derivative method for finding extremum points of multi-variable functions:
\begin{equation}
\frac{\partial F}{\partial {\xi}_1}=-8A(T)\frac{3{\xi}^4_1+3{\xi}^4_2+\frac{6}{n^2}{\xi}^2_2+\frac{3}{n^4}-18{\xi}^2_1{\xi}^2_2-18\frac{{\xi}^2_1}{n^2}}{{({\xi}^2_1+{\xi}^2_2+\frac{1}{n^2})}^4}, \label{Fxi1}
\end{equation}
\begin{equation}
\frac{\partial F}{\partial {\xi}_2}=-96A(T)\frac{{\xi}_1{\xi}_2({\xi}^2_1-{\xi}^2_2-\frac{1}{n^2})}{{({\xi}^2_1+{\xi}^2_2+\frac{1}{n^2})}^4}, \label{Fxi2}
\end{equation}
The points of extremum of the forcing function $F$ are given by the condition:
\begin{equation}
\frac{\partial F}{\partial {\xi}_1}=0;\,\frac{\partial F}{\partial {\xi}_2}=0. \label{ExtrmCond}
\end{equation}
After substituting equations (\ref{Fxi1}) and (\ref{Fxi2}) in the above condition (\ref{ExtrmCond}), we get the extremum points of $F$; which are given by
\begin{equation}
({\xi}_1,{\xi}_2)=(\pm \frac{\sqrt{3\pm 2\sqrt{2}}}{n},0)=(\pm 2.414,0)\,and\,(\pm 0.414,0). \label{ExtrmPts}
\end{equation}
Therefore, there are four extremum points of the forcing function $F$; two of which are as seen in Figure-6 at both $T=0$ and $T=1$. It should be noted here that these four extremum points of $F$, i.e. two minima and two maxima, move with the same velocity as the extremum point of $U$, i.e. one maxima. This fact can also be visualized from Figure-6, where two extremum points, i.e. one maxima and one minima, move with the same velocity. This again explains the pinned nature of analytical lump wave solution $U$ and forcing lump wave $F$. One important characteristic of our pinned lump wave solution is that even if $U$ and $F$ are pinned to each other, they do not overlap each other; that makes our pinned lump waves special. These special pinned accelerated magnetosonic lump waves can be detectable easily in the desired LEO region, which represent indirect signatures of space debris particles. This is due to the fact that these lump waves $U$ and $F$ are not overlapping each other. To further consolidate this fact, we have plotted both $U$ and $F$ against $X$ at a particular constant value of $Z$ for different times. This is as shown in Figure-7.

\begin{figure}
\includegraphics[height=7cm]{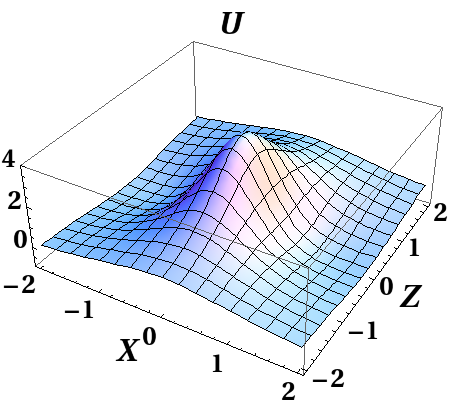}
 \ \ \ \ \includegraphics[height=7cm]{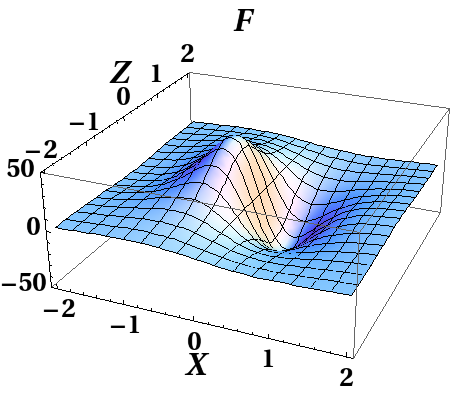}

\vspace{0.5cm}

\qquad \qquad (a) The lump wave $U$ at $T=0$
\quad \qquad \qquad \quad \qquad (b) The debris function $F$ at $T=0$
 
\vspace{1cm}

\includegraphics[height=7cm]{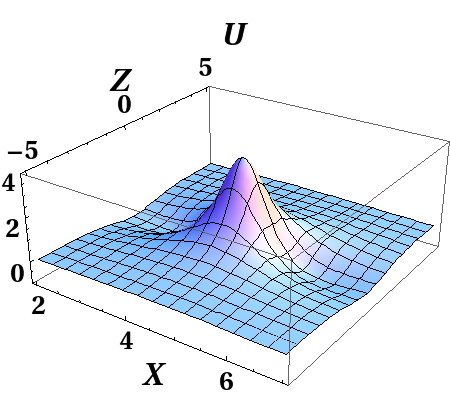}
 \ \ \ \includegraphics[height=7cm]{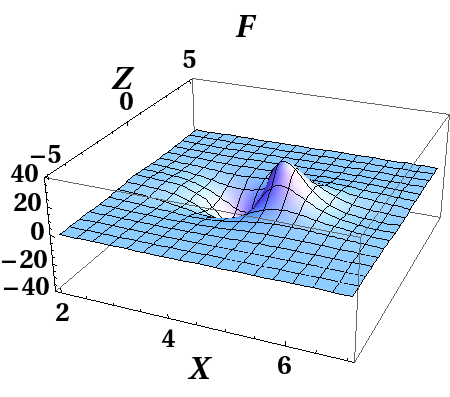}

\vspace{0.5cm}

 (c) The lump wave $U$ at $T=1$
\quad \qquad \qquad \qquad \qquad (d) The debris function $F$ at $T=1$

 \vspace{0.5cm}

 \caption{3D plots of lump solution U (\ref{PinnedLump1}) and debris field $F$ (\ref{FExpanded}) in $X-Z$ plane for $A(T)=8\cos{4T}$ and $n=1,m=0$ at two different times. We can see that the two waves move with the same velocity, hence they can be called as pinned accelerated lump waves.} 
%\vspace{0.5cm} 
\end{figure}
\begin{figure}
\includegraphics[width=16cm]{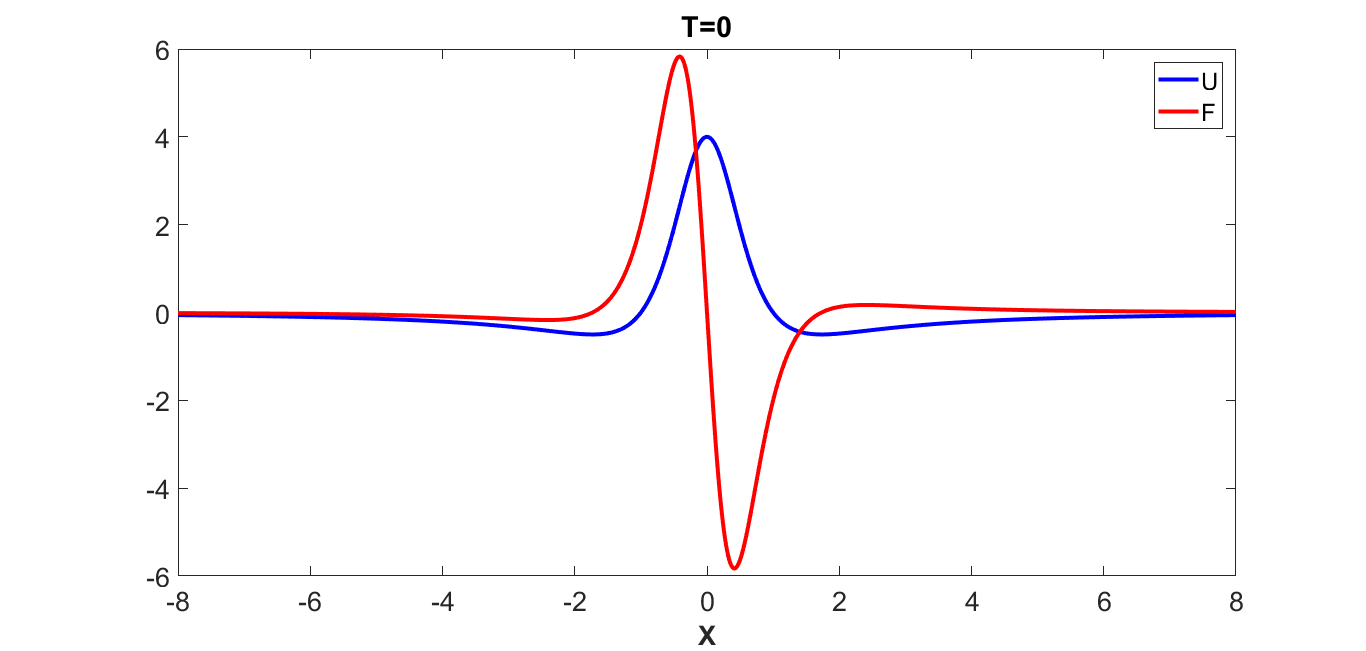}
\ \  \includegraphics[width=16cm]{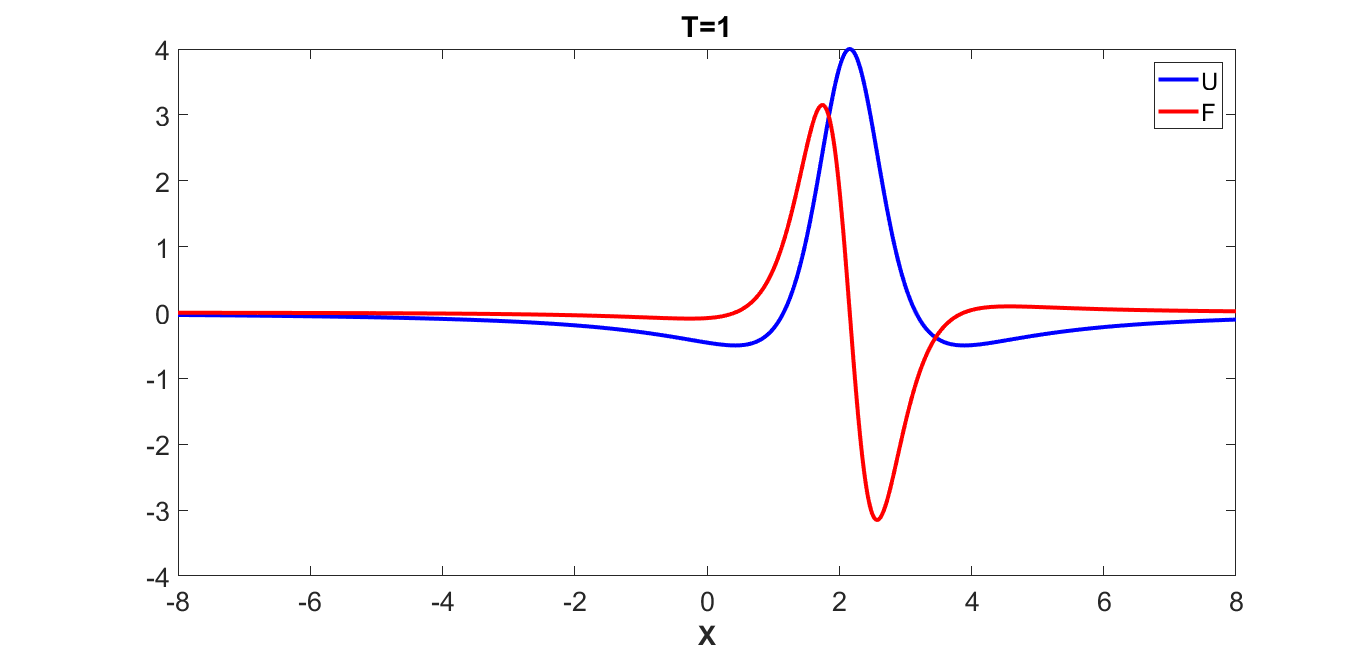}

%\vspace{0.5cm}

%\qquad \qquad(a) Pinned lump waves $U$ and $F$ at $T=0$ \qquad \qquad \qquad \qquad(b) Pinned lump waves $U$ and $F$ at $T=1$
 
%\vspace{0.5cm}

\caption{Variations of accelerated lump wave solution $U$ given by equation (\ref{PinnedLump1}) and forcing function $F$ given by equation (\ref{FExpanded}) versus $X$ for $m=0,\,n=1$ and $A(T)=cos(T)$ on $Z=0$ plane at $T=0$ and $T=1$. The plots clearly indicate that $U$ and $F$ are pinned to each other, i.e. they move with the same velocity} 
\end{figure}
 \ \ \ %\includegraphics[width=7cm]%{F1.png}

\subsection{Approximate lump wave solutions in presence of charged space debris current} \label{ApproxLump}
In previous subsections, we have derived special exact accelerated and constrained lump solutions for the magnetosonic wave having interesting analytic properties. But in spite of its novelty, it is obtained for  definite self consistent forms of $F$ as seen in (\ref{omega_for}), (\ref{Cons_lump}) and (\ref{fURel}). For other forms of $F$, the exact solvability of the fKP equation may not be possible. But on the other hand, for a weak self consistent debris field $F$, an approximate lump wave solution can be evaluated in a general way.

It is well-known that KP equation satisfies infinitely many conservation laws due to its complete integrability. These include conservation of mass, conservation of momentum, conservation of energy etc. Various conservation laws along with detailed symmetry analysis for the KP equation are explored in \cite{Anco, Zhang}.  But, in presence of an external forcing term, these quantities are not conserved in general. It is very interesting to see how these quantities change in presence of the external forcing term. In particular, we analyse how the momentum conservation relation in absence of external forcing gets modified after introduction of the forcing term. Following some mathematical simplifications from equation (\ref{FinalKPTrans}), after multiplying both sides by $U$ and integrating between limits $-\infty$ and $+\infty$ with respect to both $X$ and $Z$, we obtain the modifications in momentum conservation relation as
\begin{equation}
\frac{\partial}{\partial T}(\int_{- \infty}^{\infty} \int_{- \infty}^{\infty} U^2 dX dZ)=2\int_{- \infty}^{\infty} \int_{- \infty}^{\infty} U F dX dZ, \label{MomtConsvn}
\end{equation}
where  the solitary wave solution $U$ and its spatial derivatives are assumed to vanish when $X$ and $Z$ tend to $\pm \infty$.  It should also be noted here that the modifications in conservation relations, particularly momentum conservation relation, in presence of different types of external dissipative terms have been studied long ago by Ott and Sudan \cite{Ott} in one dimension for the case of KdV equation. Now we proceed to analyse how the lump solutions  suffer changes in dynamical behaviour due to modifications in the momentum conservation relation given by equation (\ref{MomtConsvn}) in presence of the forcing term; which represents the effects of debris objects in our system.

For this, we assume that the amplitude and velocity of the lump solution have slow time dependences as consequences of the external forcing term. Then, the standard lump solution from the forced KPI equation becomes
\begin{equation}
U=4\frac{-{[X+mZ+3(m^2-n^2(T))T]}^2+n^2(T){(Z+6mT)}^2+\frac{1}{n^2(T)}}{{\lbrace {[X+mZ+3(m^2-n^2(T))T]}^2+n^2(T){(Z+6mT)}^2+\frac{1}{n^2(T)} \rbrace}^2}, \label{DecayLump}
\end{equation}
where the maximum amplitude of lump wave is $4n^2(T)$; which propagate with velocity components $v_X=3[m^2+n^2(T)]$ and $v_Z=-6m$. This means that the parameter $n$ becomes time dependent in presence of external forcing; thereby, modifying both the amplitude and velocity of the lump wave solution. This kind of analysis for approximate lump wave solutions was also reported by Janaki et al. \cite{Janaki}. We obtain the nature of the time dependence of the parameter $n(T)$, i.e. the time dependences of maximum amplitudes and velocities of the lump wave solution $U$, for various types of localized space debris functions as given below.
\subsubsection{$F=\delta U$}
 In this subsection, we assume that the forcing term $F$ is self-consistently related to the analytical solution $U$ as $F=\delta U$ with $\delta$ as a small positive number. Substituting this and lump solution (\ref{DecayLump}) in the momentum conservation relation (\ref{MomtConsvn}), we obtain after doing the double integration
\begin{equation}
\frac{d n(T)}{dT}=2 \delta n(T). \label{Udndt}
\end{equation}
Integrating both sides of the above equation (\ref{Udndt}) from initial time $T_0$ to final time $T$, it can be easily obtained:
\begin{equation}
n(T)=n(0) exp[2 \delta (T-T_0)], \label{UApprox1}
\end{equation}
where $n(0)$ and $n(T)$ denote the values of the time dependent parameter $n$ at times $T_0$ and $T$ respectively. From equation (\ref{FrameTrans}), it is obvious that the normalized forcing function $F$ is negative whereas the normalized analytical solution $U$ is positive; this is because the nonlinear, dispersive and $2D$ coefficients $N,\,D$ and $T$ are positive whereas the external forcing coefficient $E$ is negative for the parameter space where we are analyzing the dynamics of lump waves. By considering this fact, the expression (\ref{UApprox1}) becomes
\begin{equation}
n(T)=n(0) exp[-2 \delta (T-T_0)], \label{UApprox}
\end{equation}
 Therefore, finally we get an exponentially decaying function for the amplitudes and velocities of lump waves. This is plotted in Figure-8. The analytical solution $U$ given by equation (\ref{DecayLump}), in turn, becomes 
\begin{equation}
U=4\frac{-{{\chi}_1}^2+n^2(0)exp[-4\delta(T-T_0)]{{\eta}_1}^2+n^{-2}(0)exp[4\delta(T-T_0)]}{{\lbrace {{\chi}_1}^2+n^2(0)exp[-4\delta(T-T_0)]{{\eta}_1}^2+n^{-2}(0)exp[4\delta(T-T_0)] \rbrace}^2}, \label{DecayLumpU}
\end{equation}
where we have defined two new variables for simplicity: ${\chi}_1=X+mZ+3 \lbrace m^2-n^2(0)exp[-4\delta(T-T_0)]\rbrace T$, and ${\eta}_1=Z+6mT$, and we have used equation (\ref{UApprox}) for substitution of $n(T)$. This implies that the forcing function $F=\delta U$ can be written as:
\begin{equation}
F=4\delta \frac{-{{\chi}_1}^2+n^2(0)exp[-4\delta(T-T_0)]{{\eta}_1}^2+n^{-2}(0)exp[4\delta(T-T_0)]}{{\lbrace {{\chi}_1}^2+n^2(0)exp[-4\delta(T-T_0)]{{\eta}_1}^2+n^{-2}(0)exp[4\delta(T-T_0)] \rbrace}^2}. \label{DecayLumpUF}
\end{equation}
 From equations (\ref{DecayLumpU}) and (\ref{DecayLumpUF}), it is clear that the analytical solution $U$ and forcing function $F$ differ only by a factor $\delta$, which is usually very small. Therefore, the time-dependent amplitudes of $U$ and $F$ are different; that are related by a factor of $\delta$, whereas the time-dependent velocities of both $U$ and $F$ are the same. This indicates that the analytical lump wave solution $U$ and forcing debris function $F$, which is taken to be of the form of lump wave, are pinned to each other during their resonant or self-consistent propagation in the LEO region, i.e. they move with the same velocities in spite of having different amplitudes. Again, as the velocities of both $U$ and $F$ are time-dependent, these are called approximate pinned accelerated magnetosonic lump solitons.

 In a similar way, we can get an exponentially growing function for the amplitudes and velocities of lump waves; in the case where the polarity of the forcing function $F$ is reversed, i.e. $F=-U$, in the parameter space where dynamics of lump waves is considered. It should be noted here that the growth in the amplitudes and velocities lump waves occurs at the expense of loss of energy from the physical space debris in the LEO plasma region. In earlier studies \cite{Kulikov, Mukherjee} on the dynamics of space debris in the LEO plasma region, growth in amplitudes and velocities of nonlinear solitary waves is also reported; where this growth is explained to happen at the cost of gaining energy from space debris.   
\begin{figure}[hbt!]
\centering
\includegraphics[width=10cm]{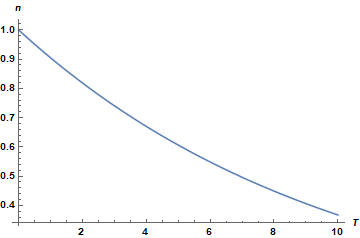}
\caption{Variation of $n(T)$ against $T$ for the case $F=\delta U$ with parameter values $\delta =0.05,\,n(0)=1$ and $T_0=0$}
\label{FU}
\end{figure}
\subsubsection{$F=\delta U_X$}
 In this subsection, we assume that the forcing term $F$ is self-consistently related to the analytical solution $U$ as $F=\delta U_X$. Substituting this and lump solution (\ref{DecayLump}) in the momentum conservation relation (\ref{MomtConsvn}), we obtain
\begin{equation}
\frac{d n(T)}{dT}=0. \label{UXdndt}
\end{equation}
Therefore, in this special case, the time dependence of $n(T)$ is frozen, and it remains constant as like in case of unforced KP equation. This indicates that momentum is conserved for this special case, and the amplitudes and velocities of the lump wave solution $U$ remain constant, i.e. $U$ is identical to the solution of unforced KPI equation.  It can also be easily
verified that for this choice of $F$, the RHS of the equation (\ref{MomtConsvn}) becomes zero after integration, due to the localized nature of $U$.
\subsubsection{$F=\delta {U_{X}}^2$}
 In this subsection, we assume that the forcing term $F$ is self-consistently related to the analytical solution $U$ as $F=\delta {U_X}^2$. Substituting this and lump solution (\ref{DecayLump}) in the momentum conservation relation (\ref{MomtConsvn}), we obtain after completing the double integration
\begin{equation}
\frac{d n(T)}{dT}=\frac{16}{35}\delta n^7(T). \label{UXSdndt}
\end{equation}
Integrating both sides of the above equation (\ref{UXSdndt}) from initial time $T_0$ to final time $T$, we get
\begin{equation}
n(T)=n(0) {[1-\frac{96}{35} \delta n^6(0)(T-T_0)]}^{-1/6}. \label{UXSApprox}
\end{equation}
Again, by taking into account the normalizations given by equation (\ref{FrameTrans}) as like previous subsections, the above expression for $n(T)$ modifies as
\begin{equation}
n(T)=n(0){[1+\frac{96}{35} \delta n^6(0)(T-T_0)]}^{-1/6}. \label{UXSApprox}
\end{equation}
This variation of $n(T)$ is shown in Figure-9. Therefore, again we get a decaying function for the amplitudes and velocities of lump waves for this special case. The explicit expression for the analytical lump wave solution $U$ with decaying amplitudes and velocities, in this case, is given by
\begin{equation}
U = 4 [\frac{-{{\chi}_2}^2+n^2(0){[1+\frac{96}{35} \delta n^6(0)(T-T_0)]}^{-1/3}{{\eta}_1}^2+n^{-2}(0){[1+\frac{96}{35} \delta n^6(0)(T-T_0)]}^{1/3}}{{\lbrace {{\chi}_2}^2+n^2(0){[1+\frac{96}{35} \delta n^6(0)(T-T_0)]}^{-1/3}{{\eta}_1}^2+ n^{-2}(0){[1+\frac{96}{35} \delta n^6(0)(T-T_0)]}^{1/3}\rbrace}^2}], \label{DecayLumpUXS}
\end{equation}
where ${\chi}_2=X+mZ+3\lbrace m^2-n^2(0){[1+\frac{96}{35} \delta n^6(0)(T-T_0)]}^{-1/3} \rbrace T$, and we have used the equation (\ref{UXSApprox}). It should be noted here that the nature of decay for the amplitudes and velocities of the lump wave solution (\ref{DecayLumpUXS}) is not exponential; rather given by a power law decay. Therefore, decay of the amplitude and velocity of the above lump soliton solution (\ref{DecayLumpUXS}) is slow as compared to that given by the lump soliton solution (\ref{DecayLumpU}) of the first subsection. If we reverse the polarity of the forcing function $F$, i.e. $F=-U$, then a growing function for the amplitudes and velocities of lump waves will be resulted. 
\begin{figure}[hbt!]
\centering
\includegraphics[width=10cm]{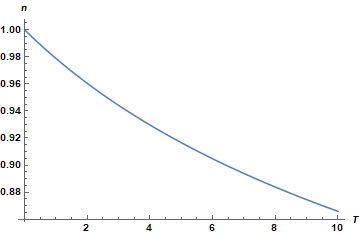}
\caption{Variation of $n(T)$ against $T$ for the case $F=\delta U^2_X$ with parameter values $\delta =0.05,\,n(0)=1$ and $T_0=0$}
\label{FUXS}
\end{figure}

In the same way, we can evaluate the approximate lump wave solutions for other weak self-consistent functional forms of forcing debris function $F$. These possible self-consistent forms of $F$ are to be chosen such that they retain lump-type structures which are appropriate for descriptions of effects of space debris particles in the LEO plasma region. This is because lump-type structures effectively behave like two-dimensional Gaussian functions; which are derived to represent debris particles by Truitt et al. \cite{TruittKP} as discussed in the previous subsection.

 \section{Discussions and possible  applications} \label{ResDis}
  In this section, we recapitulate the new findings of this article along with elaborate discussions on the possible technique of detection of space debris objects through observation of stable lump solitons. 
 \begin{enumerate}
 \item
  We have derived an exact accelerated lump wave solution as represented in equation (\ref{PinnedLump1}). The characteristic dynamical behaviour of this solution as a function of the propagation angle $\alpha$ in original coordinate system, i.e. $\xi,\,\zeta$ and $\tau$, keeping other parameters and variables constant, is shown in Figure-10. From this figure, it is obvious that the mass density ${\rho}_1$ is mostly concentrated for values of $\alpha$ which lie neither close to $0$ nor to $\pi /2$ for both slow and fast magnetosonic lump waves. This means that the distribution of mass density in the $1D$ space spanned by propagation angle $\alpha$, provided all other parameters and variables are kept constant, retains the localized nature with an approximate Gaussian shape for slow waves, and a superposition of two approximate Gaussian shapes for fast waves. 
\begin{figure}[hbt!]
\centering
\includegraphics[width=18cm]{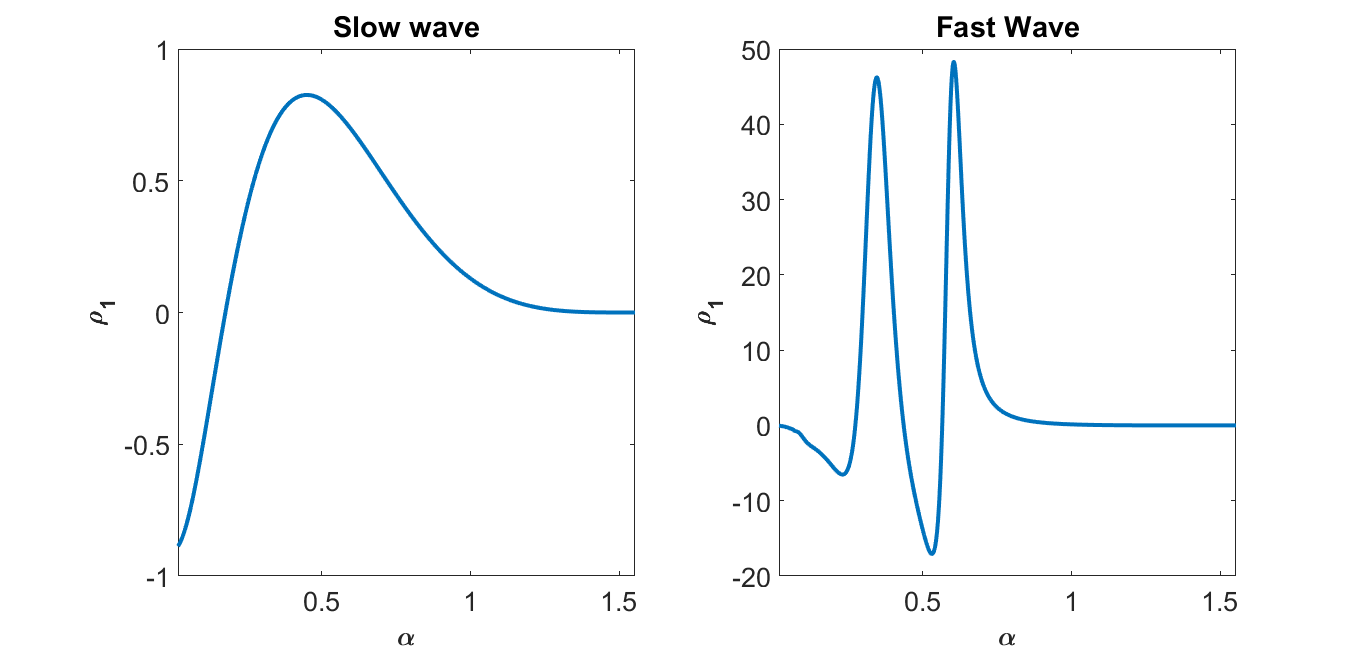}
\caption{Characteristic variation of ion mass density ${\rho}_1$, derived from exact accelerated lump wave solution \ref{PinnedLump1}, versus angle of propagation $\alpha$ for $\beta=0.8,\,m=0.1$ and $n=1$ at $\xi=1,\,\zeta=1$ and $\tau
=1$; with $\alpha$ represented in radian unit. Here $A(T)=cos(T)$ with $T=D\tau$ for slow magnetosonic lump waves whereas $T=-D\tau$ for fast magnetosonic lump waves. This figure is associated with both slow and fast magnetosonic lump waves as specified in it}  
\end{figure} 
 \item
  In the last decade, Sen et al. \cite{Sen} proposed an indirect method of detection of centimetre-sized debris objects by observation of precursor line solitons. But they neglected  effects like the time dependences of amplitudes and velocities of the forcing functions which represent real debris fields.  These forcing functions are subsequently generalized by Mukherjee et al. \cite{Mukherjee}, who concluded that changes in amplitudes and velocities of nonlinear ion acoustic solitary waves can represent indirect signatures of space debris objects. This detection technique of Mukherjee et al. is also strengthened by the earlier work of Kulikov and Zak \cite{Kulikov}, who have concluded that increase in amplitudes of nonlinear waves in the LEO region can imply the presence of debris objects. In all these works, effects of dust cloud on the solitary waves in the LEO region are neglected. These effects of dust particles are theorized in our recent work \cite{Acharya} on dust ion acoustic waves in presence of charged space debris. There we conclude the bending phenomenon of dust ion acoustic solitary waves due to charged debris motion in the LEO region in $(2+1)$ dimensions.
\item
In this work, we investigate the debris dynamics by taking into account the ambient magnetic field in the LEO region; which is caused by the Earth's magnetosphere, interplanetary magnetic field etc. A detailed theoretical analysis of space debris dynamics in the LEO region considering the surrounding magnetic field has not been reported till now as far as our knowledge goes. Recently, Kumar and Sen \cite{Kumar} have performed the numerical analysis of generation of precursor magnetosonic solitons due to space debris objects, in presence of the magnetic field in $Z$ direction, through particle in cell (PIC) simulation. They observe the bending phenomenon of magnetosonic waves in $Y$ direction when a $2D$ circular source term is taken into account. Therefore, they conclude that the presence of magnetosonic solitary waves due to debris objects is not to be uniformly felt in $Y$ direction. For the first time, we derive a forced Kadomtsev-Petviashvili equation, in this work, in presence of the ambient magnetic field in the LEO region. In our earlier investigations on space debris \cite{Mukherjee, Acharya}, we have derived forced Korteweg-de Vries (KdV) equaton and forced Kadomtsev-Petviashvili equation for ion acoustic waves and dust ion acoustic waves respectively in presence of space debris. We have not considered the penetrating magnetic field in the LEO plasma region in these earlier studies.
 \item
Following the recent work of Ruderman on Hall plasmas \cite{Ruderman}, it can be concluded that lump waves are stable solutions for both slow magnetosonic and fast magnetosonic waves. It is well-known that lump wave solutions are special kinds of rational function solutions that are localized in all directions in space whereas normal soliton solutions are exponentially localized solutions in certain directions. As per our results, the stability of lump waves come at the expense of instability of slow and fast magnetosonic solitary waves. In particular, we obtain the lump wave solution of the form (\ref{Lump_c}); when the forcing function $F$ satisfies a specific constraint condition given by equations (\ref{omega_for}) and (\ref{Cons_lump}). But the forcing function, which represent effects of space debris particles, is not very liable to satisfy such a constraint condition in realistic circumstances. We have also discovered exact as well as approximate pinned accelerated lump soliton solutions arising out of charged space debris motion in the LEO region; as explored in detail in the previous section. The magnetosonic lump wave and forcing debris function, which is taken self-consistently to be of localized forms including lump types propagate with  time-dependent velocities. Changes in both amplitudes and velocities are observed for our exact as well as approximate accelerated lump wave solutions as discussed in the previous section. Taking into account these novel results, a detection system can be envisaged for orbital charged debris particles through observations of lump solitons as well as changes in amplitudes and velocities of lump solitons. These changes in amplitudes and velocities can be detected easily through different advanced techniques; both ground-based and in situ. In addition, it should be mentioned that magnetosonic lump solitons reported in this work have widths of the order of ion inertial lengths thereby permitting large footprints in the LEO plasma region enabling their easy detection. Therefore, there are finite chances that lumps, being comparatively more stable than line solitons, can be detected more efficiently in realistic LEO plasma region. 
\item
In this theoretical investigation on magnetosonic lump waves generated due to orbital motion of charged space debris particles, we have derived a forced KP equation in $(2+1)$ dimensions and explored various analytical solutions of this equation with appropriate physical interpretations. In order to visualize space debris dynamics in the LEO region from a practical viewpoint, our work need to be extended to $(3+1)$ dimensions. In addition, the onset of multiple soliton solutions including possibility of multiple lump wave solutions, and their interaction solutions from the forced KP equation in $(3+1)$ dimensions can provide a detailed dynamical evolution of space debris particles in the LEO region. Multiple soliton solutions and interaction solutions from KP equation are explored in \cite{MaYL, Yu} as well without considering external forcing term. These interesting features of solitons apprehended to be induced by space debris is planned to be theorized in our future explorations on space debris using proper nonlinear mathematical modelling.     
  \end{enumerate}
\section{Conclusions} \label{Concl}
 We obtain a forced KP  equation in this work describing the dynamical evolution of nonlinear magnetosonic waves in presence of space debris in the LEO plasma region; which is a low density and low temperature plasma containing an enormous number of space debris objects in presence of the ambient magnetic field. Thereafter, we analyse different types of possible solutions of this nonlinear evolution equation through various analytical techniques along with their stability behavior. Lump soliton solutions retain their stability for our system in presence of charged space debris at the expense of instability of slow and fast magnetosonic waves. Exact as well as approximate pinned accelerated lump soliton solutions have been derived for the first time as a result of charged space debris motion. We propose that this lump wave solution, being localized in nature in all directions in space, can represent an indirect signature of the presence of orbital debris objects; enabling a way of indirect detection of debris objects through observations of lump solitons along with changes in their amplitudes and velocities. Thus, our work provides a detailed dynamical evolution of nonlinear magnetosonic lump waves generated in the LEO plasma region by charged space debris motion; taking into account various realistic circumstances that are not hitherto investigated by researchers working in this field as far as our knowledge goes.
 \section{Acknowledgements}
 Siba Prasad Acharya acknowledges the financial support received from Department of Atomic Energy (DAE) of Government of India, during this work through institute fellowship scheme. Abhik Mukherjee acknowledges Indian Statistical Institute, Kolkata, India for the financial support during the progress of the work.
 
\section{Declaration}
1. Funding: Department of Atomic energy, Government of India and Indian Statistical Institute, Kolkata, India

2. Conflicts of interest/Competing interests: Not applicable

3. Availability of data and material: All data are available in the article

4. Code availability: Not applicable

5. Authors' contributions: Siba Prasad Acharya (first author) is lead author; Abhik Mukherjee and M. S. Janaki are supporting authors.

%6. Ethics approval: Yes

%7. Consent to participate: Yes

%8. Consent for publication: Yes
\end{document}